\documentclass[12pt]{article}
\usepackage{amsfonts}
\usepackage{amsmath}
\usepackage{geometry}

\newtheorem{theorem}{Theorem}

\newtheorem{example}[theorem]{Example}

\input{tcilatex}
\begin{document}

\title{\textbf{Fixed Point Iteration for Estimating The Parameters of
Extreme Value Distributions}}
\author{T\textsc{ewfik} K\textsc{ernane}$^{1}\thanks{%
Corresponding author: tkernane@gmail.com}$ and Z\textsc{ohrh} A. R\textsc{%
aizah}$^{2}$ \\
$^{1}$\textit{Department of Mathematics, Faculty of Sciences}\\
$^{2}$\textit{Scientific Departments, Girls Faculty of Education}\\
\textit{King Khaled University}\\
\textit{Abha, Saudi Arabia}}
\maketitle

\begin{abstract}
Maximum likelihood estimations for the parameters of extreme value
distributions are discussed in this paper using fixed point iteration. The
commonly used numerical approach for adressing this problem is the
Newton-Raphson approach which requires differentiation unlike the fixed
point iteration which is also easier to implement. Graphical approaches are
also usualy proposed in the literature. We prove that these reduce in fact
to the fixed point solution proposed in this paper.
\end{abstract}

\textbf{2000 AMS\ Classification}: 62F10; 62N02.

\section{Introduction}

Extreme value distributions are largely used in applied engineering and
environmental problems (see the book by Coles \cite{Col}). Parameter
estimation is the first step in the statistical analysis of parametric
probability distributions. The most widely used approach is the popular
maximum likelihood estimation (MLE), but usually, as for extreme value
distributions, the solution is not analytic and must be then approached by
numerical techniques. The commonly used one is the Newton-Raphson algorithm
to determine the maximum likelihood estimates of the parameters. To employ
the algorithm, the second derivatives of the log-likelihood are required.
Sometimes the calculations of the derivatives based on the progressively
Type II censored samples for example are complicated (see \cite{Bala2}). To
avoid such computation, we propose to use the fixed point iteration
algorithm instead. In this paper we prove that generally MLE of extreme
value distributions could be expressed in a fixed point iteration form which
is easier to implement and does not require differentiation. Graphical
techniques were also proposed as alternatives (see Balakrishnan and Kateri 
\cite{Bala1} and Dodson \cite{Dod}) which have the disadvantage of using
visualization to detect the solution from graphics. We prove that these
graphical solutions reduce in fact to fixed points of a suitable iteration
forms.

In the following section, we propose fixed point iterations for estimating
the parameters of the Gumbel and Weibull distributions from complete data.
In section 3, we extend the procedure to censored samples (simple and
progressive Type I and Type II) using the Type I least extreme values
distribution from which estimations for Gumbel and Weibull distributions can
be deduced from suitable transformations. Finally, in section 4 we
illustrate the proposed approach using examples quoted from the literature.

\section{Maximum Likelihood Estimations for Complete Data}

\subsection{Type I extreme value distribution}

The Type I extreme value distribution which is also called Gumbel
distribution function is defined as:%
\begin{equation*}
F(x)=\exp \left\{ -\exp \left[ -\frac{1}{\sigma }\left( x-\mu \right) \right]
\right\} ,
\end{equation*}%
for $x\in \mathbb{R},$ and it has the probability density%
\begin{equation*}
f(x)=\frac{1}{\sigma }\exp \left[ -\frac{1}{\sigma }\left( x-\mu \right) %
\right] \exp \left\{ -\exp \left[ -\frac{1}{\sigma }\left( x-\mu \right) %
\right] \right\} ,
\end{equation*}%
where $\sigma >0$ and $\mu \in \mathbb{R}.$ Let $\underline{x}%
=(x_{1},...,x_{n})^{\prime }$ be a complete sample from the Type I extreme
value distribution.

The maximum likelihood estimator for $\sigma $ is known to be the solution
of the following equation

\begin{equation}
\sigma =\frac{\sum_{i=1}^{n}x_{i}}{n}-\frac{\sum_{i=1}^{n}x_{i}\exp (-\frac{%
x_{i}}{\sigma })}{\sum_{i=1}^{n}\exp (-\frac{x_{i}}{\sigma })}.
\label{sigmahat}
\end{equation}%
Denote the right hand side of (\ref{sigmahat}) by $g(\sigma ;\underline{x}).$
Then the equation (\ref{sigmahat}) is in fact in a fixed point form%
\begin{equation}
\sigma =g(\sigma ;\underline{x}).  \label{fixp1}
\end{equation}%
Unlike the Newton-Raphson method, which is the commonly used nonlinear
numerical method for solving MLE of the extreme value distribution (see
Cohen \cite{Coh1} p. 143), the fixed point approach does not require
differentiation and then is more easier to use. Uniqueness of the solution
of (\ref{sigmahat}) can be proved using graphical techniques (see
Balakrishnan and Kateri \cite{Bala1}). Indeed, the left hand side $\sigma $
is monotone increasing on $\sigma .$ We then have to show that $g(\sigma ;%
\underline{x})$ is monotone decreasing on $\sigma .$%
\begin{equation*}
\frac{\partial g(\sigma ;\underline{x})}{\partial \sigma }=\frac{%
\sum_{i=1}^{n}\frac{x_{i}}{\sigma ^{2}}\exp (-\frac{x_{i}}{\sigma }%
)\sum_{i=1}^{n}x_{i}\exp (-\frac{x_{i}}{\sigma })-\sum_{i=1}^{n}\frac{%
x_{i}^{2}}{\sigma ^{2}}\exp (-\frac{x_{i}}{\sigma })\left(
\sum_{i=1}^{n}\exp (-\frac{x_{i}}{\sigma })\right) }{\left(
\sum_{i=1}^{n}\exp (-\frac{x_{i}}{\sigma })\right) ^{2}}
\end{equation*}%
It remains then to prove that%
\begin{equation*}
g^{\ast }(\sigma ;\underline{x})=\sum_{i=1}^{n}x_{i}\exp \left( -\frac{x_{i}%
}{\sigma }\right) \sum_{i=1}^{n}x_{i}\exp \left( -\frac{x_{i}}{\sigma }%
\right) -\sum_{i=1}^{n}x_{i}^{2}\exp \left( -\frac{x_{i}}{\sigma }\right)
\left( \sum_{i=1}^{n}\exp \left( -\frac{x_{i}}{\sigma }\right) \right) \leq
0.
\end{equation*}

Setting $a_{i}=x_{i}\exp \left( -\frac{x_{i}}{2\sigma }\right) $ and $%
b_{i}=\exp \left( -\frac{x_{i}}{2\sigma }\right) ,$ $i=1,...,n$, $g^{\ast
}(\sigma ;\underline{x})$ becomes%
\begin{equation*}
g^{\ast }(\sigma ;\underline{x})=\left( \sum_{i=1}^{n}a_{i}b_{i}\right)
^{2}-\sum_{i=1}^{n}a_{i}^{2}\sum_{i=1}^{n}b_{i}^{2}
\end{equation*}%
and $g^{\ast }(\sigma ;\underline{x})\leq 0$ by the Cauchy-Schwarz
inequality. The last result can be deduced from the result of Balakrishnan
and Kateri \cite{Bala1} from transformation between the Weibull and Type I
extreme value distribution. It should be noted that $\lim_{\sigma
\rightarrow \infty }g(\sigma ;\underline{x})=0$ and%
\begin{equation*}
\lim_{\sigma \rightarrow 0^{+}}g(\sigma ;\underline{x})=\frac{%
\sum_{i=1}^{n}x_{(i)}}{n}-x_{(1)}=\frac{1}{n}\sum\limits_{i=2}^{n}\left(
x_{(i)}-x_{(1)}\right) .
\end{equation*}%
We deduce then that $0\leq \sigma \leq \frac{\sum_{i=1}^{n}x_{(i)}}{n}%
-x_{(1)}$ and $0\leq g(\sigma ;\underline{x})\leq \frac{\sum_{i=1}^{n}x_{(i)}%
}{n}-x_{(1)}$ which also guarantees the existence of a solution (see Lemma
3.4.1 of \cite{Atk}). After obtaining a solution for $\sigma $ we deduce the
MLE of $\mu $ from%
\begin{equation}
\widehat{\mu }=\sigma \left[ \ln \frac{n}{\sum_{i=1}^{n}\exp (-\frac{x_{i}}{%
\sigma })}\right] .  \label{mu1}
\end{equation}

\begin{example}
Consider the data about annual wind-speed maxima in km/h from 1947 to 1984
at Vancouver quoted from the software Xtremes 4.1 \cite{Xtr}, provided with
the book by Reiss and Thomas \cite{Rei}, stored in the file em-cwind.dat
(the source is \cite{Gen}). By fitting a Gumbel distribution to the data,
they provide estimates for $\sigma $ and $\mu $ as $\widehat{\sigma }=8.3$
and $\widehat{\mu }=60.3.$ The fixed point iteration (\ref{fixp1}) leads to
the solution $\widehat{\sigma }=8.2891$ and from (\ref{mu1}) we obtain $%
\widehat{\mu }=60.3504.$
\end{example}

\subsection{Weibull distribution}

For the two parameters Weibull distribution $\mathcal{W}(\theta ,\beta )$
with pdf%
\begin{equation*}
f\left( x;\theta ,\beta \right) =\frac{\beta }{\theta ^{\beta }}x^{\beta
-1}\exp \left[ -\left( x/\theta \right) ^{\beta }\right] ,\text{ }x>0,\text{ 
}\theta ,\beta >0,
\end{equation*}%
which is the Type II extreme value distribution it can be deduced from \cite%
{Bala1} that 
\begin{equation}
\beta =g_{w}(\beta ;\underline{x}),  \label{weib-fix}
\end{equation}%
where $g_{w}(\beta ;\underline{x})$ is given by%
\begin{equation*}
g_{w}(\beta ;\underline{x})=\left[ \frac{\sum\limits_{i=1}^{n}x_{i}^{\beta
}\ln x_{i}}{\sum\limits_{i=1}^{n}x_{i}^{\beta }}-\frac{1}{n}%
\sum\limits_{i=1}^{n}\ln x_{i}\right] ^{-1}.
\end{equation*}%
It has been proved in \cite{Bala1} that $\left( g_{w}(\beta ;\underline{x}%
)\right) ^{-1}$ is a monotone increasing function then $g_{w}(\beta ;%
\underline{x})$ is monotone decreasing which insures existence and
uniqueness of the solution of the fixed point iteration (\ref{weib-fix}).
The MLE of the parameter $\theta $ is deduced from%
\begin{equation*}
\widehat{\theta }=\left( \frac{1}{n}\sum\limits_{i=1}^{n}x_{i}^{\beta
}\right) ^{1/\beta }.
\end{equation*}

\section{Estimation for Censored Data}

\subsection{Singly right censored samples}

In the case of simple Type-II censored data, let $r$ ($1<r<n$) denote the
number of observed lifetimes and $x=(x_{(1)},x_{(2)},...,x_{(r)})$ the
ordered observed lifetimes from the Type I least extreme value distribution
with probability density%
\begin{equation*}
f(x)=\frac{1}{\sigma }\exp \left[ \frac{1}{\sigma }\left( x-\mu \right) %
\right] \exp \left\{ -\exp \left[ \frac{1}{\sigma }\left( x-\mu \right) %
\right] \right\} .
\end{equation*}%
If $X$ is a random variable from a Type I greatest extreme value
distribution with location parameter $\mu $ and shape parameter $\sigma $
then $-X$ follows a Type I least extreme value distribution with location
parameter $-\mu $ and shape parameter $\sigma $ \cite{Coh1}. But if we have
a Type-II rigth censored data $x=(x_{(1)},x_{(2)},...,x_{(r)})$ from the
Type I least extreme value distribution then $%
y=(-x_{(r)},-x_{(r-1)},...,-x_{(1)})$ will be a Type-II left censored data
from the corresponding Type I greatest extreme value distribution. The MLE
of $\sigma $ is given by the following fixed point iteration expression%
\begin{equation}
\sigma =g(\sigma ;\underline{x}),  \label{sig-type2}
\end{equation}%
where now%
\begin{equation}
g(\sigma ;\underline{x})=\frac{\sum_{i=1}^{r}x_{(i)}\exp \left( -\frac{%
x_{(i)}}{\sigma }\right) +(n-r)x_{(r)}\exp \left( -\frac{x_{(r)}}{\sigma }%
\right) }{\sum_{i=1}^{r}\exp \left( -\frac{x_{(i)}}{\sigma }\right)
+(n-r)\exp \left( -\frac{x_{(r)}}{\sigma }\right) }-\frac{%
\sum_{i=1}^{r}x_{(i)}}{r}.  \label{gsig}
\end{equation}%
It can be easily proved that $g(\sigma ;\underline{x})$ of (\ref{sig-type2})
is a monotone decreasing function in $\sigma $ exactly along the lines of
section 2.1 by taking%
\begin{eqnarray*}
a_{i} &=&I\left( 1\leq i\leq r\right) x_{(i)}\exp \left( -\frac{x_{(i)}}{%
2\sigma }\right) +I\left( r+1\leq i\leq n\right) x_{(r)}\exp \left( -\frac{%
x_{(r)}}{2\sigma }\right) \\
b_{i} &=&I\left( 1\leq i\leq r\right) \exp \left( -\frac{x_{(i)}}{2\sigma }%
\right) +I\left( r+1\leq i\leq n\right) \exp \left( -\frac{x_{(r)}}{2\sigma }%
\right) ,
\end{eqnarray*}%
which guarantees existence and uniqueness of a fixed point for $\sigma .$
The parameter $\mu $ is then obtained from%
\begin{equation*}
\widehat{\mu }=\sigma \left[ \ln \frac{\sum_{i=1}^{r}\exp \left( -\frac{%
x_{(i)}}{\sigma }\right) +(n-r)\exp \left( -\frac{x_{(r)}}{\sigma }\right) }{%
r}\right] .
\end{equation*}%
If $y_{\left( 1\right) },...,y_{\left( r\right) }$ designate the $r$
smallest observations in a random sample of size $n$ from a two parameter
Weibull distribution $\mathcal{W}\left( \theta ,\beta \right) $ then $%
x_{\left( 1\right) },...,x_{\left( r\right) }$ where $x_{\left( i\right)
}=\ln y_{\left( i\right) }$ will designate equivalent observations in a
sample from the Type I distribution of smallest extremes \cite{Coh1}. The
MLEs $\widehat{\theta }$ and $\widehat{\beta }$ of the Weibull distribution
will then be deduced from the relations $\widehat{\theta }=\exp \widehat{\mu 
}$ and $\widehat{\beta }=1/\widehat{\sigma }.$ Meanwhile, fixed point
iterations hold also in this case for the Weibull distribution from the
relation%
\begin{equation}
\beta =g_{w}(\beta ;\underline{x}),  \label{gwcens1}
\end{equation}%
where now%
\begin{equation}
g_{w}(\beta ;\underline{x})^{-1}=\frac{\sum\limits_{i=1}^{r}x_{\left(
i\right) }^{\beta }\ln x_{\left( i\right) }+\left( n-r\right) x_{\left(
r\right) }^{\beta }\ln x_{\left( r\right) }}{\sum\limits_{i=1}^{r}x_{\left(
i\right) }^{\beta }+\left( n-r\right) x_{\left( r\right) }^{\beta }}-\frac{1%
}{r}\sum\limits_{i=1}^{r}\ln x_{\left( i\right) }.  \label{gwsig}
\end{equation}%
From \cite{Bala1} it has been proved that $g_{w}(\beta ;\underline{x})^{-1}$
is monotone increasing in $\beta $ then $g_{w}(\beta ;\underline{x})$ in (%
\ref{gwcens1}) is monotone decreasing which insures existence and uniqueness
of the fixed point\ $\widehat{\beta }$ of the iteration (\ref{gwcens1}).

The MLE of $\theta $ is deduced from%
\begin{equation}
\widehat{\theta }=\left[ \frac{1}{r}\left\{
\sum\limits_{i=1}^{r}x_{(i)}^{\beta }+\left( n-r\right) x_{\left( r\right)
}^{\beta }\right\} \right] ^{1/\beta }.  \label{theta1}
\end{equation}

For Type-I censoring, it suffices to replace the term $x_{(r)}$ in the $%
(n-r)x_{(r)}\exp \left( -x_{(r)}/\sigma \right) $ and $(n-r)\exp \left(
-x_{(r)}/\sigma \right) $ terms of relation (\ref{gsig}) by the
pre-specified time of testing $T$ and in the terms $\left( n-r\right)
x_{\left( r\right) }^{\beta }\ln x_{\left( r\right) }$ and $\left(
n-r\right) x_{\left( r\right) }^{\beta }$ in the relation (\ref{gwsig}).

\begin{example}
In this example we quote a Type-II censored data from \cite{Dod}. These data
are about testing twenty identical grinders, with the test ending at time
152.7. In this period, twelve grinders failed. The observed failure times
are presented in the following table%
\begin{eqnarray*}
&&\text{\textbf{Table 1}} \\
&&\text{Type-II censored failure data from Dodson (2006)} \\
&&%
\begin{tabular}{llllllllllll}
\hline
12.5 & 24.4 & 58.2 & 68.0 & 69.1 & 95.5 & 96.6 & 97.0 & 114.2 & 123.2 & 125.6
& 152.7 \\ \hline
\end{tabular}%
\end{eqnarray*}
By assuming a Weibull distribution the graphical approach of \cite{Bala1}
leads to solutions $\widehat{\beta }=1.647$ and $\widehat{\theta }=162.223$.
Using the fixed point iteration (\ref{gwcens1}) we obtain the fixed point
solution $\widehat{\beta }=1.6467$ and from (\ref{theta1}) $\widehat{\theta }%
=162.223$. This shows that the graphical approach proposed in \cite{Bala1}
reduces in fact to a fixed point iteration.
\end{example}

\subsection{Progressively censored samples}

Consider a progressively Type II censored sample from a Type I least extreme
values distribution with $R_{j},$ $j=1,...,r$ the number of censored items
at failure time $j$. We have the following fixed point iteration for the MLE
of $\sigma $%
\begin{equation}
\sigma =g(\sigma ;\underline{x}),  \label{sig-prog}
\end{equation}%
where%
\begin{equation*}
g(\sigma ;\underline{x})=\frac{\sum_{i=1}^{r}\left( R_{i}+1\right) x_{\left(
i\right) }\exp \left( \frac{x_{(i)}}{\sigma }\right) }{\sum_{i=1}^{r}\left(
R_{i}+1\right) \exp \left( \frac{x_{(i)}}{\sigma }\right) }-\frac{1}{r}%
\sum\limits_{i=1}^{r}x_{\left( i\right) }.
\end{equation*}%
It can be proved using similar arguments of the preceding section that $%
g(\sigma ;\underline{x})$ is a decreasing function on $\sigma $ which
insures existence and uniqueness of a fixed point for $\sigma .$ The MLE of $%
\mu $ is then deduced from%
\begin{equation}
\mu =\sigma \ln \left[ \frac{1}{r}\sum_{i=1}^{r}\left( R_{i}+1\right) \exp
\left( \frac{x_{(i)}}{\sigma }\right) \right] .  \label{mu-prog}
\end{equation}

For the case of a Weibull distribution with progressively Type II censored
sample, we obtain from \cite{Bala1} that there exists a unique fixed point
solution $\widehat{\beta }$ of the following equation 
\begin{equation*}
\beta =g_{w}(\beta ;\underline{x}),
\end{equation*}%
where%
\begin{equation}
g_{w}(\beta ;\underline{x})=\left[ \frac{\sum\limits_{i=1}^{r}\left(
R_{i}+1\right) x_{\left( i\right) }^{\beta }\ln x_{\left( i\right) }}{%
\sum\limits_{i=1}^{r}\left( R_{i}+1\right) x_{\left( i\right) }^{\beta }}-%
\frac{1}{r}\sum\limits_{i=1}^{r}\ln x_{\left( i\right) }\right] ^{-1}.
\label{sig-progtype2}
\end{equation}%
We have for the MLE\ of $\theta $%
\begin{equation}
\widehat{\theta }=\left( \frac{1}{r}\sum\limits_{i=1}^{r}\left(
R_{i}+1\right) x_{(i)}^{\beta }\right) ^{1/\beta }.  \label{theta2}
\end{equation}

\begin{example}
Consider the progressive Type-II censored data analysed by Viveros and
Balakrishnan \cite{Viv} and given in the following table%
\begin{eqnarray*}
&&\text{\textbf{Table 2}} \\
&&\text{Progressive Type-II censored failure data from Viveros and
Balakrishnan (1994) } \\
&&%
\begin{tabular}{lllllllll}
\hline
$i$ & $1$ & $2$ & $3$ & $4$ & $5$ & $6$ & $7$ & $8$ \\ \hline
$x_{(i)}$ & $-1.6608$ & $-0.2485$ & $-0.0409$ & $0.2700$ & $1.0224$ & $%
1.5789 $ & $1.8718$ & $1.9947$ \\ 
$R_{i}$ & $0$ & $0$ & $3$ & $0$ & $3$ & $0$ & $0$ & $5$ \\ \hline
\end{tabular}%
\end{eqnarray*}%
Fitting a Type-I least extreme values distribution to these data and using
the Newton-Raphson numerical method they obtain $\widehat{\sigma }=1.026$
and $\widehat{\mu }=2.222$. Using fixed point iteration (\ref{sig-prog}) we
obtain the solution $\widehat{\sigma }=1.0264$ and using (\ref{mu-prog}) we
obtain $\widehat{\mu }=2.222.$ In order to compare the convergence rate of
the fixed point iteration with that of the Newton--Raphson method and the EM
algorithm used in \cite{Bala2}, same initial value $\sigma _{0}=0.7912$ (and 
$\mu _{0}=1.4127$ for the Newton--Raphson and EM algorithm methods) were
used and the level of accuracy was fixed at $5\times 10^{-5}$. The
Newton--Raphson method used in \cite{Viv} took $37$ iterations and the EM
algorithm took $151$ while the fixed point iteration took $12$ iterations to
converge to the same values.
\end{example}

\end{document}